\documentclass[aps,prl,reprint,showpacs,showkeys,superscriptaddress,preprintnumbers,groupedaddress]{revtex4-1}
\usepackage{amsmath,amssymb,bm,mathrsfs}
\usepackage{srcltx}
\usepackage{dsfont}
\usepackage{epsfig}
\usepackage{slashed}
\usepackage{bbold}
\usepackage{psfrag}
\usepackage{xcolor}
\PassOptionsToPackage{caption=false}{subfig}
\usepackage{subfig}
\usepackage{xfrac}
\usepackage{multirow}
\usepackage{booktabs}
\usepackage[colorlinks=true,linkcolor=blue,citecolor=blue,urlcolor=violet]{hyperref}

\newcommand{\be}{\begin{equation}}
\newcommand{\ee}{\end{equation}}
\newcommand{\bea}{\begin{eqnarray}}
\newcommand{\eea}{\end{eqnarray}}

\def\({\left(}
\def\){\right)}

\usepackage{soul}

\begin{document}

\title{Cannibal Dark Matter}

\author{Duccio Pappadopulo}
\email{duccio.pappadopulo@gmail.com}
\affiliation{
Center for Cosmology and Particle Physics, \\
Department of Physics, New York University, New York, NY 10003, USA.
}

\author{Joshua T. Ruderman}
\email{ruderman@nyu.edu}
\affiliation{
Center for Cosmology and Particle Physics, \\
Department of Physics, New York University, New York, NY 10003, USA.
} 

\author{Gabriele Trevisan}
\email{gabriele.trevisan@nyu.edu}
\affiliation{
Center for Cosmology and Particle Physics, \\
Department of Physics, New York University, New York, NY 10003, USA.
}

\begin{abstract}
A thermally decoupled hidden sector of particles, with a mass gap, generically enters a phase of cannibalism in the early Universe. The Standard Model sector becomes exponentially colder than the hidden sector. We propose the Cannibal Dark Matter framework, where dark matter  resides in a cannibalizing sector with a relic density set by 2-to-2 annihilations. Observable signals of Cannibal Dark Matter include a boosted rate for indirect detection, new relativistic degrees of freedom, and warm dark matter.
\end{abstract}

\maketitle

\noindent {\bf  Motivations.---}The unknown particle origin of Dark Matter (DM) is one of the most important problems in particle physics.  A compelling possibility is that DM is a thermal relic with abundance determined by the decoupling of its annihilations in the early Universe.

Direct detection experiments, which search for DM scattering against atomic nuclei, are making rapid progress~\cite{Cushman:2013zza,Akerib:2015rjg,Aprile:2015uzo}.  Over the next decade, direct detection will explore most of the remaining parameter space where DM scattering produces more events than background neutrinos~\cite{Billard:2013qya}.  DM will be discovered, or else powerful limits will motivate the exploration of scenarios where DM interacts feebly with the Standard Model (SM).

If DM has sizable couplings to SM particles, then its temperature in the early Universe tracks the photon temperature.  Alternatively, if the rate of DM interactions with SM particles is slower than the Hubble expansion, DM evolves with a different temperature than photons.  In the latter case, we can define a dark sector with temperature $T_d \ne T_\gamma$, which contains DM and any other particles in kinetic equilibrium with DM\@.

The DM relic density depends crucially on the properties of the dark sector when DM annihilations decouple.  There are two classes of dark sectors, depending on the mass of the Lightest Dark sector Particle (LDP), $m_{\rm LDP}$:
\begin{enumerate}
\item relativistic LDP: $m_{\rm LDP} < T_{f}$
\item non-relativistic dark sector:  $m_{\rm LDP} > T_{f}$,
\end{enumerate}
where $T_{f}$ is the dark sector temperature when DM annihilations freeze-out. We note that the LDP may not be the DM\@.
Existing studies with hidden sector DM abundance set by 2-to-2 annihilations assume the presence of a relativistic LDP (see for example Refs.~\cite{Strassler:2006im,Pospelov:2007mp,Feng:2008ya,Feng:2008mu,ArkaniHamed:2008qn,Kaplan:2009ag,Cheung:2010gj,D'Agnolo:2015koa}).  

In the second class of models, under the generic requirement that number changing interactions are active, the hidden sector undergoes a phase of {\it cannibalism}.  As we review below, the properties of cannibalism are determined by the requirement that the dark sector and SM separately preserve their comoving entropy densities.  This leads to different scalings of the temperature versus the scale factor $a$,
\begin{equation}\label{eq.temp_dep}
T_{\gamma} \propto 1/a \qquad {\rm and}\qquad T_{d} \sim 1 / \log a,
\end{equation}
where $T_{\gamma}$ ($T_{d}$) represents the temperature of the SM (cannibalizing) sector. The hidden sector temperature stays almost constant as the Universe expands as number changing interactions efficiently convert the rest mass of non-relativistic particles into kinetic energy \cite{Carlson:1992fn}. The different dependence on the scale factor implies that the SM particles are exponentially colder than the cannibalizing sector, as a function of the hidden sector temperature.
Ref.~\cite{Carlson:1992fn} was the first to propose (and name) cannibalism.  In Ref.~\cite{Carlson:1992fn} and followup studies with cannibalism~\cite{deLaix:1995vi,Yamanaka:2014pva,Bernal:2015ova,Bernal:2015xba,Kuflik:2015isi,Soni:2016gzf}, it is assumed that DM annihilates through a 3-to-2 (or 4-to-2) process.  Refs.~\cite{Hochberg:2014dra,Hochberg:2014kqa,Bernal:2015bla} consider models  where DM experiences 3-to-2 annihilation, but with DM in thermal equilibrium with radiation such that cannibalism does not occur.

Previous studies of cannibalism assume that the DM relic density comes from number changing interactions, such as 3-to-2 or 4-to-2 annihilations~\cite{deLaix:1995vi,Yamanaka:2014pva,Bernal:2015ova,Bernal:2015xba,Kuflik:2015isi,Soni:2016gzf}.  We stress that cannibalism is a generic feature of any hidden sector with a mass gap, and is therefore of broader relevance that models where the DM itself undergoes number changing annihilations.  We  initiate the study of a  new class of DM models, where DM resides in a cannibalizing sector with relic density determined by 2-to-2 annihilations.  We refer to this class of models as {\it Cannibal Dark Matter}. In general, cannibal dark matter is realized in hidden sectors that contain both metastable states, $\phi$, undergoing number changing interactions such as $\phi\phi\phi\to\phi\phi$, and a stable DM candidate, $\chi$, annihilating into the metastable states through 2-to-2 annihilations, $\chi\chi\to\phi\phi$.

The exponential cooling of the SM, during cannibalism, has dramatic implications for DM phenomenology.  DM must have a larger annihilation rate than conventional scenarios, in order to reproduce the observed relic density.  Therefore, Cannibal DM predicts boosted rates for indirect detection.  The Universe is exponentially older at DM freeze-out, implying less redshifting between DM decoupling and the start of structure formation and therefore the possibility of warm DM\@.  Cannibal DM can also lead to new relativistic degrees of freedom,  leaving imprints on the Cosmic Microwave Background (CMB).

The remainder of this letter is organized as follows.  We begin by discussing the physics of cannibalism.  Then we include DM in the cannibalistic sector and compute its relic density.  We then overview the phenomenology of our model, including indirect detection and cosmological signatures.

\begin{figure}[!!!t]
\begin{center}
\includegraphics[width=0.49 \textwidth]{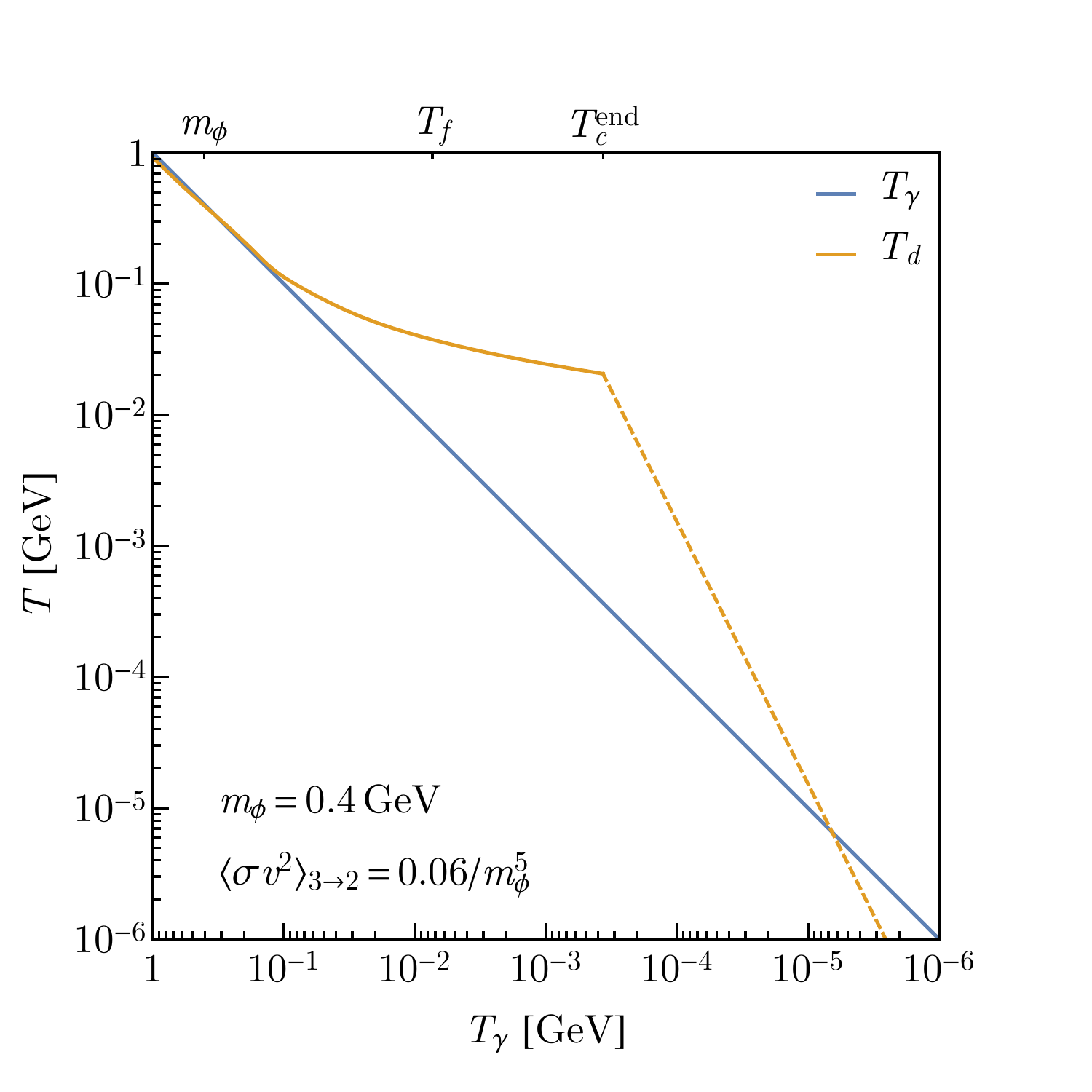}
\end{center}
\vspace{-.3cm}
\caption{\small {\em Temperature of the dark sector, $T_d$, as a function of the SM temperature, $T_\gamma$. For $T_d<m_\phi$, the SM becomes exponentially colder than the cannibalizing sector. After the end of cannibalism, $T_d < T_c^{end}$, the dark sector temperature redshifts as $a^{-2}$ due to adiabatic expansion.
}}
\label{fig.temp_rho}
\end{figure}

\vspace{.3cm}
\noindent {\bf Cannibalism.---}Cannibalism requires the following conditions:
\begin{enumerate}  \setlength\itemsep{0em}
\item The dark sector is kinetically decoupled from the SM sector.
\item The dark sector has a mass gap.
\item The dark sector remains in chemical equilibrium, through number changing interactions, at temperatures below the mass of the LDP\@.
\end{enumerate}

As a simple example, we consider a  dark sector with a real scalar LDP and generic interactions,
\begin{equation} \label{eq:Vphi}
V_\phi =  \frac{m_{\phi}^2}{2} \phi^2 + \frac{A}{3!}\,  \phi^3 + \frac{\lambda}{4!} \, \phi^4 .
\end{equation}
Cannibalism begins when $T_d$ drops below $m_\phi$, and chemical equilibrium  is maintained through $\phi \phi \phi \rightarrow \phi \phi$ annihilations. For reference we quote the thermally averaged $3\to2$ cross section obtained from Eq.~\ref{eq:Vphi} assuming $A=\sqrt{3 \lambda}m_\phi$: $\langle \sigma v^2\rangle\approx 0.04\,  \lambda^3/m_\phi^5$.

 In the following, we assume that $\phi$ is metastable and eventually decays to either SM states or dark radiation.  In order for cannibalism to occur, $\phi$ must be out of kinetic equilibrium with its decay products and therefore have lifetime longer than Hubble, $\tau_\phi > H^{-1}$, when $T_d = m_\phi$.  For decays to SM, we focus on $\phi \rightarrow \gamma \gamma$.  For the dark radiation case, we assume that $\phi$ decays to a light species that is kinetically decoupled from $\phi$ and begins with zero abundance (for example $\phi \rightarrow \gamma' \gamma'$, where $\gamma'$ is a light hidden photon).

Since the dark sector and the SM are kinetically decoupled, they have different temperatures and entropies. Assuming for simplicity that there are no entropy injections, 
 the comoving entropies of the two sectors are separately conserved. This implies that the ratio of SM to dark entropy densities is fixed,
\begin{equation}
\begin{split}
\xi \equiv \frac{s_{\rm SM}}{s_{d}}.
\end{split}
\end{equation}
If the two sectors were in thermal contact in the past, $\xi$ is the ratio of the sum of degrees of freedom within each sector~\cite{Feng:2008ya}. However, $\xi$ can be much larger if the two sectors reheat to asymmetric temperatures and remain thermally decoupled.

During cannibalism, $T_d$ and $s_d$ are related,
\begin{equation}
\begin{split}
s_{d}\equiv \frac{2\pi^2}{45}g^d_{*s}\,T_d^3\simeq \frac{m_\phi^{3}}{(2\pi)^{3/2}}x_\phi^{-1/2}e^{-x_\phi},
\end{split}
\end{equation}
where $x_\phi\equiv m_\phi/T_d$ and we assume, for simplicity, that $\phi$ dominates the hidden sector entropy.
Conservation of entropy within each sector implies that the SM becomes exponentially colder than the dark sector  (see Fig.~\ref{fig.temp_rho}),
\begin{equation}\label{temperatures}
\frac{T_{\gamma}}{T_d}\approx 0.5\,\xi^{1/3}g_*^{-1/3}x_\phi^{5/6}e^{-x_\phi/3},
\end{equation}
where $g_*$ is the effective number of relativistic degrees of freedom of the SM\@. Since $T_\gamma\sim 1/a$, Eq.~(\ref{temperatures}) leads to Eq.~(\ref{eq.temp_dep}).
The energy density in the dark sector is approximately dominated by the LDP and decreases as
\begin{equation}
\begin{split}
\rho_{d}\sim\frac{1}{a^3 \log a},
\end{split}
\end{equation}
in striking contrast to the energy density of non-relativistic matter in thermal equilibrium with a relativistic plasma, which decreases as an exponential function of the scale factor.
The dark sector starts to drive the expansion of the Universe when its energy density become equal to that of the SM, $T^E_{\gamma}/T^E_d=4/(3\xi)$, corresponding to $x^E_\phi\approx -2.8+\log (g_*^{-1}\xi^4) +2.5 \log x^E_\phi $.

Cannibalism ends when the dark sector drops out of chemical equilibrium.  This can happen for one of two reasons, (1) $\phi \phi \phi \rightarrow \phi \phi$ annihilations decouple, or (2) the Universe becomes older than the $\phi$ lifetime, and $\phi$ decays away. The second possibility requires that $ \Gamma_\phi \sim H$ when $T_d$ is between $m_\phi$ and the temperature that 3-to-2 processes decouple.   This may seem to require a coincidence of scales between the $\phi$ lifetime and Hubble.  However,   the scale factor (and therefore the age of the Universe) changes by an exponentially large amount during cannibalism,  as a function of the dark sector temperature.  Therefore, option (2) is generic and is realized for an exponentially large range of values for $\Gamma_\phi$~\cite{footnote3}. 

\begin{figure}[bt]
\begin{center}
\includegraphics[width=0.49 \textwidth]{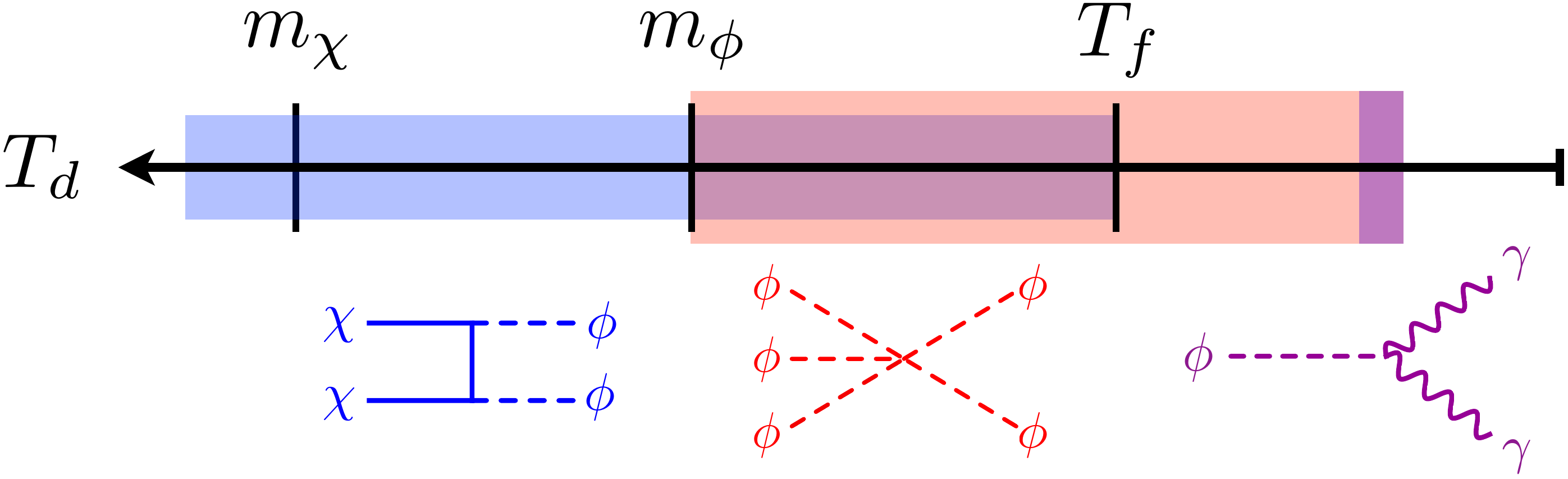}
\end{center}
\vspace{-.3cm}
\caption{\small {\em Cannibal DM has three stages. (1) DM annihilations,  $\chi \chi \rightarrow \phi \phi$, are in equilibrium and the LDP, $\phi$, is relativistic.  (2) Cannibalism begins when $\phi$ becomes non-relatavistic, and then DM annihilations freeze-out, at temperature $T_f$, during cannibalism.  (3) Cannibalism ends when $\phi$ decays away or $\phi \phi \phi \rightarrow \phi \phi$ annihilations decouple.}}
\label{fig:schema}
\end{figure}

\vspace{.3cm}
\noindent {\bf Freeze-out of Cannibal DM.---}We now include DM in the cannibalizing sector.  We consider a simple example where DM is a Majorana fermion, $\chi$, that has a Yukawa interaction with $\phi$,
\be\label{eq.ToyModel}
\mathcal L \supset - V_\phi - \left( \frac{m_\chi}{2} \chi^2 + \frac{y}{2}\,  \phi   \chi^2  + \mathrm{h.c.} \right).
\ee
$V_\phi$ is defined in Eq.~(\ref{eq:Vphi}), $m_{\chi} > m_\phi$ and $y$ is a complex number. The relic abundance of $\chi$ is determined by the freeze-out of 2-to-2 annihilations, $\chi\chi\to \phi\phi$. This cross section is $s$-wave if either Im$\,yA\neq 0$ or Im$\,y^2\neq 0$. With purely imaginary $y$ and $m_\phi=0$ this cross section reads $\langle\sigma v\rangle=\frac{A^2 |y|^2}{1024 \pi m_\chi^4}$. We assume that DM annihilations decouple during cannibalism~\cite{footnote1}.   Therefore, DM freeze-out occurs after $\phi$ becomes non-relativistic, but before $\phi \phi \phi \rightarrow \phi \phi$ annihilations decouple and before $\phi$ decays (we leave the study of different orderings of these events for future work~\cite{FPRT}).  The cosmological stages of our scenario are depicted in Fig.~\ref{fig:schema}.

\begin{figure}[!!!t]
\begin{center}
\includegraphics[width=0.49 \textwidth]{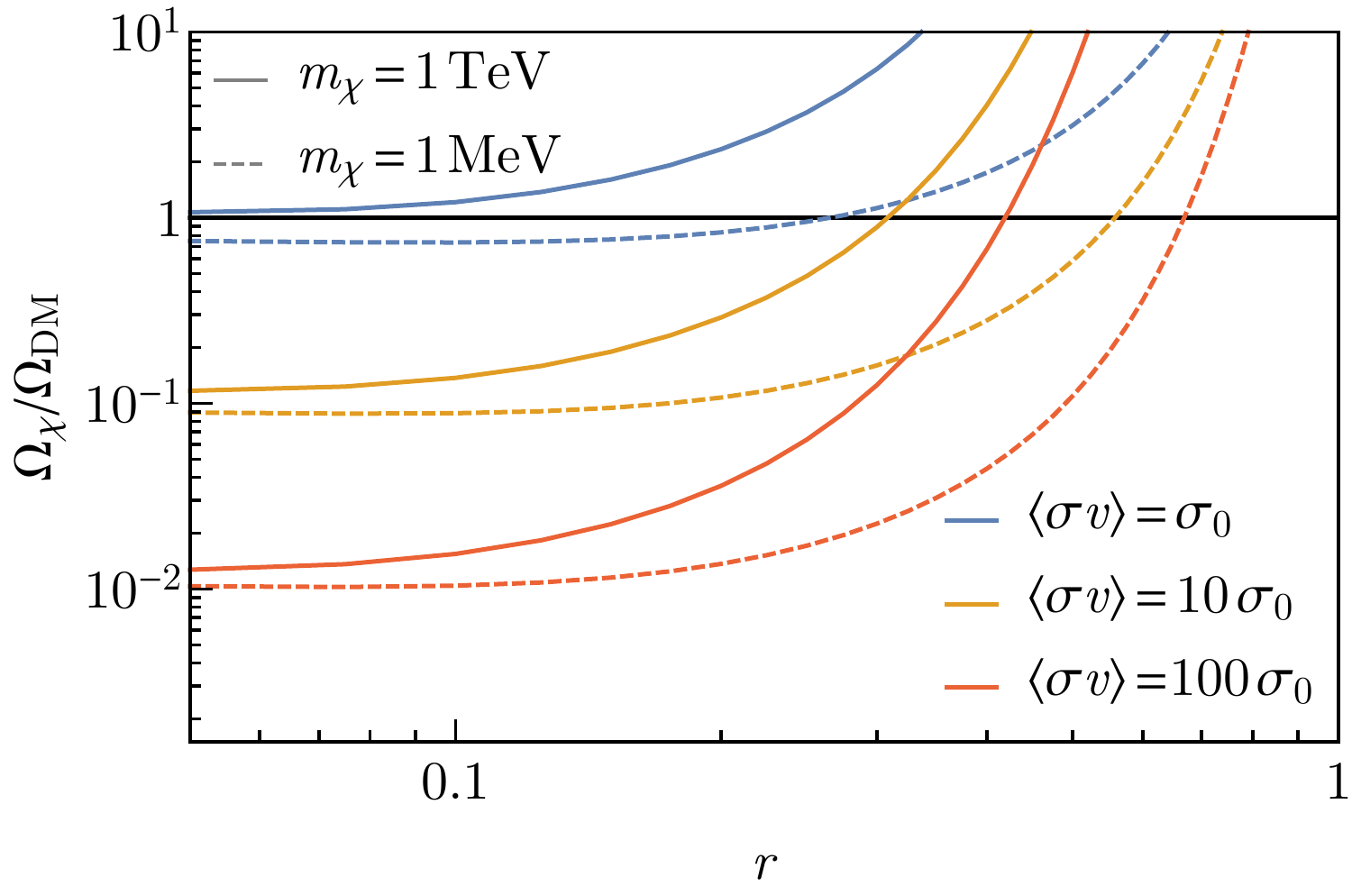}
\end{center}
\vspace{-.3cm}
\caption{\small {\em Relic density versus $r \equiv m_\phi / m_\chi$  for $m_\chi=(1\,{\rm MeV},1\,{\rm TeV})$ and $\langle\sigma v\rangle=(1,10,100)\, \sigma_0$, where $\sigma_0=3\times10^{-26}{\rm{cm}}^3{\rm{s}}^{-1}$.  The relic density grows exponentially with $r$, as implied by Eqs.~(\ref{temperatures}), (\ref{eq:yield}), and~ (\ref{eqxf}).
}}
\label{fig.omeaga_vs_r}
\end{figure}

Under the assumption that $\chi\phi\to\chi\phi$ and $\phi\phi\phi\to\phi\phi$ are in  equilibrium, the evolution of the number density of $\chi$ is described by a single Boltzmann equation, 
\begin{equation} \label{eq:boltzmann}
\frac{dY_\chi}{d\log x}=-\frac{\kappa(x)s_d\langle\sigma v\rangle}{H}(Y_\chi^2-Y^{\textrm{eq}\, 2}_\chi),
\end{equation}
where $Y_\chi \equiv n_\chi/s_d$, $x\equiv m_\chi/T_d$, and $\kappa(x)\equiv (1-1/3\, d\log g^d_{*s}/d\log x)$.  The $\chi$ relic density is given by $\Omega_\chi / \Omega_{\rm DM} = m_\chi Y_\chi / (0.4~\textrm{eV}\,\xi)$, where $\Omega_{\rm DM} h^2 \approx 0.12$ corresponds to the observed DM density~\cite{Ade:2015xua}.   A numerical solution is shown in Fig.~\ref{fig.omeaga_vs_r}.

The Boltzmann equation can be solved analytically in the sudden freeze-out approximation, $n_{\chi}^{\textrm{eq}}(x_f)\langle\sigma v\rangle= H$.  There are two regimes depending on whether the SM or $\phi$ dominate the energy density of the Universe when DM annihilations decouple,
\be\label{eq:yield}
\frac{\Omega_{\chi}}{\Omega_{\rm DM}}\approx 
\begin{cases} 
      0.3\, \frac{x_f}{g_*^{1/2}} \frac{\sigma_0}{\langle\sigma v\rangle} \frac{T_d}{T_\gamma}\times \frac{1}{D} & (\rho_{\rm SM} > \rho_d) \\
   0.3\,  \frac{x_f}{g_*^{1/2}}  \frac{\sigma_0}{\langle\sigma v\rangle} \frac{T_d^{3/2}}{\xi^{1/2} T_\gamma^{3/2}}\times\frac{1}{D}& (\rho_{\rm SM} < \rho_d)  \\
   \end{cases}
\ee
where all quantities are evaluated at $x=x_f$ and $\sigma_0=3\times10^{-26}{\textrm{cm}}^3{\textrm{s}}^{-1}$.
$D$ is a dilution factor that accounts for the entropy which is injected in the SM plasma when $\phi$ decays. $D=1$ if $\phi$ energy density does not dominate at the time it decays; otherwise we evaluate $D$ using a sudden decay approximation \cite{Kolb:1990vq}, $D=T_\gamma^E/T_{RH}$, where $T_{RH}\approx 0.8\, g_*^{-1/4}\Gamma_\phi^{1/2}M_P^{1/2}$ is the  temperature that the SM is reheated to after $\phi$ decays~\cite{footnote2}.  

According to Eq.~(\ref{temperatures}), $T_d/T_\gamma \sim e^{r x_f/3}$, which is naturally very large during cannibalism, requiring a boosted annihilation cross section compared to conventional scenarios.  Eq.~(\ref{eq:yield}) depends on the temperature DM annihilations decouple, $x_f = m_\chi / T_f$, which in the sudden freeze-out approximation is,
\be\label{eqxf}
x_f\approx \delta^{-1} \log [\,h(r)\, m_\chi M_P\langle\sigma v\rangle]\,,
\ee
where $\delta=1-\tfrac{2}{3}r$ or  $1-\tfrac{1}{3}r$ for SM or $\phi$ domination, respectively, and
\be
h(r)  \approx 
\begin{cases}
	0.3\, g_*^{\frac{1}{6}}\xi^{-\frac{2}{3}} r^{-\frac{5}{3}}(1-\tfrac{2}{3}r)^{\frac{7}{6}}  & (\rho_{\rm SM} > \rho_d) \\
   0.2\, r^{-\frac{5}{4}}(1-\tfrac{1}{2}r)^{\frac{3}{4}}  &  (\rho_{\rm SM} < \rho_d)   \\
   \end{cases}
\ee
Eqs.~(\ref{temperatures}), (\ref{eq:yield}), and  (\ref{eqxf}) imply that the final DM abundance is exponentially sensitive to $r$, as shown in Fig.~\ref{fig.omeaga_vs_r}.

\vspace{.3cm}

\begin{figure*}[tb]
\centering
\includegraphics[width=1\linewidth]{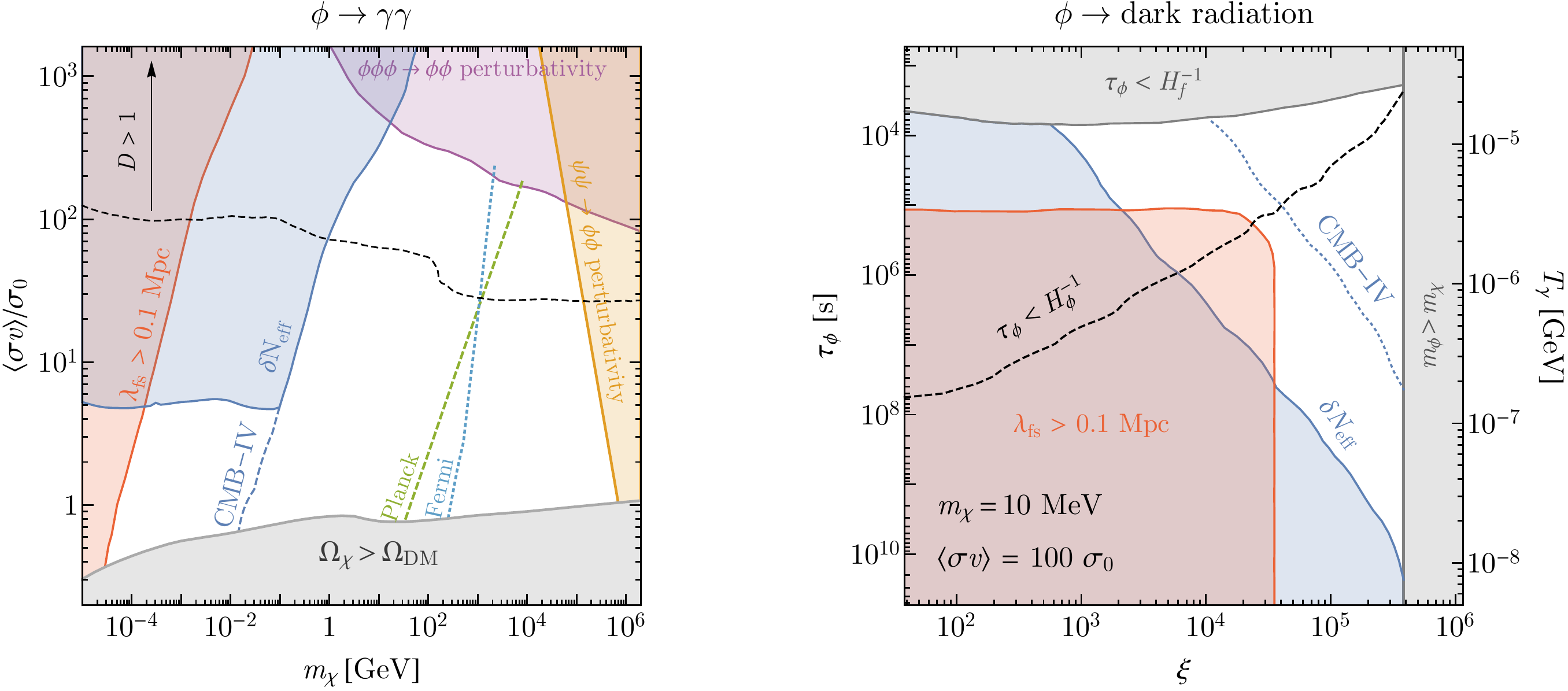}
\vspace{-.3cm}
\caption{
\label{fig:pheno}
\em
Constraints on the model of Eq.~(\ref{eq.ToyModel}), fixing $r = m_\chi / m_\phi$ such that $\Omega_\chi = \Omega_{\rm DM}$.
In the red  region, the DM free streaming length is inconsistent with measurements of the Lyman-alpha forest: $\lambda_{\textrm{fs}} > 0.1$~Mpc~\cite{Viel:2013apy,Baur:2015jsy}. 
 \textbf{Left:}~We take $\phi$ to decay to photons, $\phi \rightarrow \gamma \gamma$, with $\tau_\phi\approx H_f^{-1}=H^{-1}(T_f)$ (to show the full parameter space), and $\xi = s_{\rm SM} / s_d \approx 39$.     
DM annihilations, $\chi \chi \rightarrow  \phi \phi \rightarrow 4 \gamma$, are constrained by Planck~\cite{Ade:2015xua,Slatyer:2015jla,Elor:2015bho} and Fermi~\cite{Ackermann:2015zua,Elor:2015bho}.  The displayed bounds correspond to $s$-wave annihilations, and are superseded by other limits for $p$-wave annihilations.  Perturbativity constraints are shown: $\sigma_{3\to2}<(4\pi)^5/m_\phi^5$ and $\sigma_{2\to2}<(4\pi)^3/m_\chi^2$.     Above the black dashed line, $\phi$ dominates the energy density of the Universe before decaying, producing entropy in the SM sector and diluting the DM abundance by $D > 1$.   \textbf{Right:} Constraints when $\phi$ decays to dark radiation, as a function of $\xi$ and $\tau_\phi$ (or of $T_\gamma$ when $\phi$ decays). The  blue shaded region (dashed line) corresponds to the limit~\cite{Ade:2015xua} (reach~\cite{Wu:2014hta}) on enhanced $N_{\rm eff}$ due to dark radiation.  Above (below) the black dashed line, cannibalism ends when $\phi$ decays ($\phi \phi \phi \rightarrow \phi \phi$ decouples).  In the upper gray region, cannibalism ends before DM annihilations freeze-out.}
\label{fig.Money}
\end{figure*}

\noindent {\bf Phenomenology.---}  Cannibal DM is not presently testable by direct detection, because the SM and hidden sector are kinetically decoupled, implying a small cross section.  However,  Cannibal DM predicts rich signals in indirect detection and cosmology, driven by the boosted DM annihilation rate, the (exponentially large) age of the Universe at DM freeze-out, and the decay of relic LDPs.

When $\phi$ decays, its energy density can modify the number of relativistic degrees of freedom, which is constrained by the CMB: $N_{\rm eff}=3.15~\pm~0.23$ ($1\sigma$)~\cite{Ade:2015xua}.
If $\phi$ decays to photons after neutrino decoupling, the photons are heated relative to the neutrinos, lowering $N_{\textrm{eff}}$. Alternatively,  if $\phi$ decays to dark radiation the resulting energy density  increases $N_{\textrm{eff}}$. These constraints  are shown in Fig.~\ref{fig:pheno}, fixing $r= m_\chi / m_\phi$ at each point such that $\Omega_\chi = \Omega_{\rm DM}$. In the left panel, we assume $\phi$ decays to photons and $\xi=\xi_0\approx 39$, which is the entropy ratio if the SM and the hidden sector of Eq.~(\ref{eq.ToyModel}) were in thermal contact above the weak scale. In the right panel, we assume that $\phi$ decays to dark radiation and we allow $\xi$ to vary.
In this  case, the constraint on  $\delta N_{\textrm{eff}}$ excludes the possibility of $\phi$ ever dominating the expansion of the Universe. In Fig.~\ref{fig:pheno}, we also show the projected sensitivity on $N_{\textrm{eff}}$ coming from CMB Stage-IV experiments, $\delta N_{\rm eff} = 0.03$ ($2 \sigma$)~\cite{Wu:2014hta}.  We have verified that $\delta N_{\rm eff}$ constraints from Big Bang Nucleosynthesis (BBN)~\cite{Cooke:2013cba,Aver:2013wba} are subdominant to the above constraints.

When $\phi$ decays to photons, there are indirect detection constraints on the process $\chi \chi \rightarrow \phi \phi \rightarrow 4 \gamma$.  Energy injection at the recombination epoch can distort the CMB as measured by Planck~\cite{Ade:2015xua,Slatyer:2015jla,Elor:2015bho}. Fermi observations of dwarf galaxies~\cite{Ackermann:2015zua} bound DM annihilations at the present epoch~\cite{Elor:2015bho}. These constraints are shown in the left panel of Fig.~\ref{fig:pheno}. The Planck and Fermi limits are shown for $s$-wave $\chi\chi\to\phi\phi$ annihilations and they become irrelevant if these annihilation are $p$-wave~\cite{Essig:2013goa}. This can be easily achieved in the model of Eq.~\ref{eq.ToyModel} by Im$\,y=0$.

An important constraint on Cannibal DM follows from the exponentially long time it takes DM to freeze-out. While the average velocity of Cannibal DM at freeze-out is of the same order as conventional scenarios, $v_f\sim x_f^{-1/2}$, the SM is exponentially colder at freeze-out, Eq.~(\ref{temperatures}).  Between kinetic decoupling and matter-radiation equality, DM free-streaming damps density perturbations, resulting in a cut-off in the matter power-spectrum on scales smaller than the free-streaming length, $\lambda_{\textrm{fs}}$. 
This is typically a negligible effect for standard scenarios with DM mass above a few keV since $\lambda_{\textrm{fs}}\propto T^{-3/2}_{\textrm{k}}$, with $T_{\textrm{k}}$ the temperature of kinetic decoupling of DM (however heavy and warm DM is possible if DM is produced from late decays of a heavier state~\cite{Cembranos:2005us,Kaplinghat:2005sy}).  In Fig.~\ref{fig:pheno}, kinetic decoupling occurs after the end of cannibalism and therefore $T_{\rm k}$ is exponentially smaller than $m_\chi$, making free-streaming relevant when $m_\chi \gg$~1 keV\@. The strongest constraint on the free-streaming length, shown in Fig.~\ref{fig:pheno}, comes from observations of the Lyman-alpha forest: $\lambda_{\rm fs}\lesssim0.1{\rm\,Mpc}$~($2\sigma$)~\cite{Viel:2013apy,Baur:2015jsy}.  In order to evaluate the bound we use the approximate expression
\be\label{freestreaming}
\lambda_{\textrm{fs}}=\int_{t_{\rm k}}^{t_{\rm eq}}\frac{v(t)}{a(t)}dt\approx 125\,{\textrm{Mpc}}\, v_{\textrm{k}}\frac{\log (1.3\,T_{\textrm k}^{\textrm{eV}})}{T_{\textrm k}^{\textrm{eV}}},
\ee
where $v_{\textrm{k}}$ is the average DM velocity at kinetic decoupling and $T_{\textrm k}^{\textrm{eV}}$ is the relative SM plasma temperature in eV\@. Eq.~\ref{freestreaming} assumes the universe is radiation dominated between decoupling and equality, which is always true in our case.
The right panel of Fig.~\ref{fig:pheno} shows that the free-streaming length can be the strongest constraint for a DM particle of thermal origin and mass of 10~MeV\@.

\vspace{.3cm}
\noindent {\bf Outlook.---}In this letter, we have introduced the Cannibal DM framework (Fig.~\ref{fig:schema}). DM resides in a cannibalistic sector and has relic density set by the freeze-out of 2-to-2 annihilations.  Cannibalism has a significant impact on DM phenomenology (Fig.~\ref{fig:pheno}), including a boosted annihilation rate, warm DM, and new relativistic degrees of freedom.

We conclude by noting that  cannibalism  has unique implications for early Universe cosmology that deserve further study.  During cannibalism, the Hubble constant decreases as an exponential function of the hidden sector temperature, changing the clock by which all hidden sector processes are measured.  It would be interesting to exploit this property to explore further implications of cannibalism, such as new DM production mechanisms and the possibility that baryogenesis is connected to cannibalism.

\vspace{.3cm}
\begin{acknowledgements}
\noindent {\bf \em Acknowledgements.---}We thank Raffaele Tito D'Agnolo, Marco Farina, Martin Schmaltz, Kris Sigurdson, Matteo Viel, Neal Weiner, and Cheuk Yin Yu for helpful discussions. D.~P. and G.~T. are supported by the James Arthur Postdoctoral Fellowship. 
\end{acknowledgements}



\begin{thebibliography}{0}
\bibitem{footnote3}
For instance, requiring that $\phi$ decays while non-relativistic but before it dominates the energy density, the $\phi$ lifetime scales as $\xi^{8/3}$  and spans 5 orders of magnitudes for $\xi=39$.
\end{thebibliography}
\end{document}